# Hardware Acceleration for Open Radio Access Networks: A Contemporary Overview

Lopamudra Kundu, Xingqin Lin, Elena Agostini and Vikrama Ditya

NVIDIA

Emails: {lkundu, xingqinl, eagostini, vditya}@nvidia.com

*Abstract*—Radio access networks (RAN) are going through a paradigm shift towards interoperable, intelligent, software-defined, and cloud-native open RAN solutions. A key challenge towards the adoption and deployment of open RAN at scale is performance, since commercial RAN must meet stringent requirements in terms of capacity, data rate, latency, coverage, energy efficiency, reliability, and security, among others. To overcome the performance challenge, it is critical to leverage the power of hardware acceleration to offload compute-heavy RAN workloads to specialized hardware devices to enable accelerated compute for open RAN deployments. In this article, we provide a state-of-the-art overview of hardware acceleration for open RAN in the fifth generation (5G) wireless networks, by reviewing RAN architectural evolution, presenting the role of hardware acceleration in RAN, and introducing the latest O-RAN Alliance's development of acceleration abstraction layer. We also present a practical implementation of inline hardware acceleration for open RAN layer 1 processing and identify several areas for future exploration towards the sixth generation (6G) wireless networks.

## I. INTRODUCTION

The evolution of global wireless technology to the fifth generation (5G) and beyond enables a new kind of network that not only enhances broadband paradigm, but also aims to make ubiquitous connectivity a reality. To unleash the full potential of 5G, new trends are emerging towards making network deployments more dynamic, agile, and energy efficient. Radio access network (RAN) is arguably the most complex and compute-intensive component of the wireless network infrastructure. In order to meet the ever-increasing demand for network traffic, operators are looking into fundamentally new ways to optimize their RAN deployment.

In recent years, there has been a growing trend towards RAN disaggregation, challenging the traditional RAN infrastructures that are largely siloed, monolithic, and proprietary [1]. Widely known as 'open RAN', this emerging RAN deployment approach is based on the principles of modular components and open interfaces. By utilizing commercial-off-the-shelf (COTS) hardware, virtualized network functions, and decoupled software, open RAN brings with it many benefits such as scalability, flexibility, cost efficiency, and rapid innovation [2][2]. To continue the momentum of open RAN transformation trend, two key criteria are critical - scalability and compute.

One of the main challenges of open RAN adoption at scale is to achieve high, end-to-end network performance and in turn, meet the unprecedented demand of RAN processing for 5G and beyond [4]. With key features including wide channel bandwidths, support for low latency, sophisticated channel coding schemes (e.g., low-density parity check (LDPC) code, polar code), and advanced antenna technologies with massive multiple-input multiple-output (MIMO) antenna arrays, the compute burden of 5G RAN has grown significantly compared to its predecessors [5]. Therefore, using only general-purpose processors (GPPs) to realize 5G RAN can lead to unacceptable power consumption and in turn, impractical network deployment option. To overcome this challenge, it is critical to leverage the power of hardware acceleration to offload compute-heavy RAN workloads to specialized hardware devices such as graphics processing unit (GPU), data processing unit (DPU), application-specific integrated circuit (ASIC), field programmable gate array (FPGA), digital signal processor (DSP) or system-on-chip (SoC), enabling accelerated compute for 5G RAN.

The importance of hardware acceleration for open RAN is being increasingly recognized by the wireless and networking communities. An FPGA-based LDPC accelerator for 5G open RAN was introduced in [6]. Similarly, the work in [6] investigated the feasibility and efficiency of cloud native 5G RAN acceleration by offloading LDPC decoding to FPGA. A comprehensive survey on non-binary LDPC codes was presented in [8], which compared ASIC, FPGA, and GPU based architectures. A summary of the open RAN architecture developed by the O-RAN Alliance was presented in [9], but the work did not provide sufficient coverage on the roles of hardware acceleration in open RAN.

The objective of this article is to provide a state-of-the-art overview of hardware acceleration in open RAN, addressing the gap in the existing literature. To that end, the rest of the paper is outlined as follows. Section II provides an overview of RAN architectural evolution, paving the way towards open RAN. Section III illustrates the role of hardware acceleration in RAN. Its manifestation into the open RAN domain with hardware-software decoupling through acceleration abstraction layer is explained in Section IV. In Section V, we deep dive into a use case of hardware acceleration in open RAN. Finally, Section VI wraps up this paper with conclusions and future directions.

## II. EVOLUTION OF RAN ARCHITECTURE

A RAN consists of a group of base stations (BSs) that provide wireless access to mobile users. The BS is composed of two major parts: baseband unit (BBU) and remote radio unit (RRU). The BBU typically handles most of the digital signal processing for wireless transmission and reception, while the RRU is mainly responsible for the analog and radio frequency (RF) signal processing. Hosting both the BBU and the RRU at an antenna site is known as a distributed RAN (D-RAN) architecture. In D-RAN, the transport between the antennas site and the core network is referred to as backhaul. D-RAN is a commonly used architecture in the existing cellular networks. It is particularly well suited in areas with low population density and scattered users. However, D-RAN is inflexible when it comes to managing capacity needs at the BSs. Most of the BSs in D-RAN use dedicated and overprovisioned hardware to provide capacity for peak demand, resulting in underutilized resources most of the time.

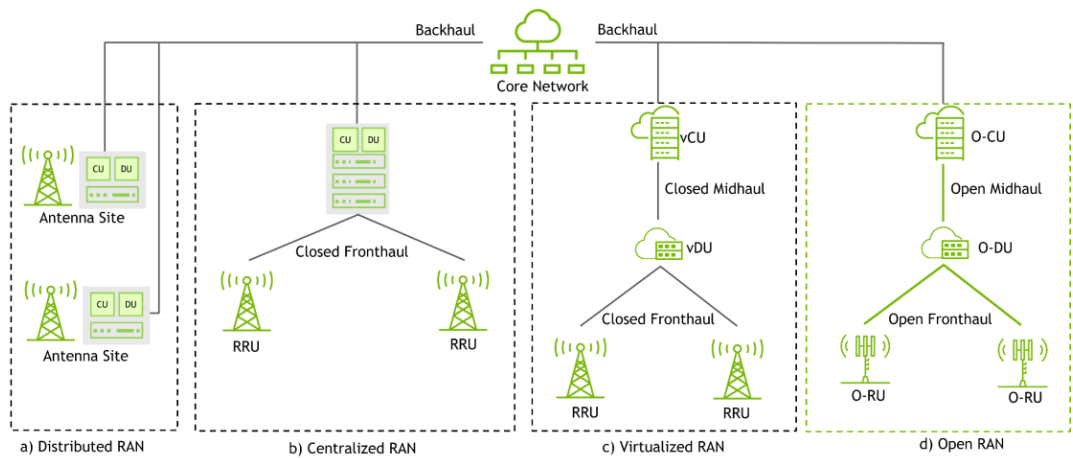

**Figure 1 An illustration of RAN architecture options: D-RAN, C-RAN, vRAN and Open RAN**

In a centralized RAN (C-RAN) architecture, the BBUs are aggregated and located in a central site or data center, while the RRUs stay at the antenna sites. The transport between the BBUs and the RRUs is referred to as fronthaul. Compared to backhaul, fronthaul in general has more stringent requirements (e.g., shorter latency and wider bandwidth). With BBU pooling, the compute, storage, and network resources can be shared among multiple BBUs in an elastic and scalable way. C-RAN is well suited for use in urban areas with high population density. Figure 1 provides illustrations of centralized and distributed RAN architecture options. For both C-RAN and D-RAN, it is possible to split a 5G node B (gNB) into a centralized unit (CU) and one or more distributed units (DUs).

Traditional RAN deployments in the 4G Long-Term Evolution (LTE) era with D-RAN and C-RAN topologies were predominantly brownfield, with tightly coupled hardware and software solutions from a single vendor, integrated through proprietary interfaces. This legacy 'black box' approach, however, is largely inflexible in terms of reconfigurability and scalability - the essential traits for the 5G network. The first step towards 5G RAN evolution is realized through virtualization, transforming the legacy RAN into a virtualized RAN (vRAN) [2], depicted in Figure 1c. Breaking away from the monolithic architecture of the conventional RAN , a vRAN decouples RAN application software from the underlying hardware platform using the principles of network functions virtualization [11]. In vRAN, RAN applications within CU(s) and DU(s) are deployed as virtual network functions (VNFs) or cloud-native network functions (CNFs) running on COTS servers instead of proprietary, purpose-built hardware as in legacy RAN.

As the momentum of 5G RAN transformation continues, vRAN further evolves into open RAN - a concept built-upon the principles of open interfaces and virtualization. Even though vRAN introduces a certain degree of disaggregation between software and hardware, it still remains a single vendor, closed-box solution, with proprietary interfaces interconnecting its components. Open RAN further disaggregates vRAN into a multi-vendor, software-defined, white-box solution running on COTS servers, with open and standardized interfaces and protocols between its open nodes including O-RAN CU (O-CU), O-RAN DU (O-DU), O-RAN radio unit (O-RU), as illustrated in Figure 1d. With open RAN comes a paradigm shift in the way mobile network operators (MNOs) fundamentally plan, deploy, and operate their networks. Breaking the barrier of vendor lock-in, open RAN enables abundant flexibility for MNOs to build their RANs by integrating best-of-the-breed components from a diverse supply chain through interoperable interfaces, to meet scaling and compute demands of 5G RAN. By the virtue of openness and virtualization, open RAN is poised to democratize the vendor ecosystem, lower the barrier of entry for new vendors, and foster competition and innovation. At the forefront of this transformative journey is O-RAN Alliance [1], a global consortium formed by leading MNOs and hundreds of RAN vendors from around the world. Technical specifications developed by O-RAN provide the guiding principles for harmonizing the open RAN transformation across the industry to ensure alignment and compatibility.

Across this entire deployment landscape for 5G RAN, hardware accelerators (HWAs) play a crucial role in meeting the unprecedented computing demand of the wireless network. In the next section, we will see how hardware acceleration in RAN is evolving as the deployment of 5G is gearing up around the globe.

## III. HARDWARE ACCELERATION IN RAN

Within the RAN protocol stack, the processing of gNB-DU scales with number of connections, transmission time interval (TTI), bandwidth, and antenna branches. Usually, the lower the layer in the RAN protocol stack, the higher the demand on the compute. In particular, the need for PHY layer processing within gNB-DU increases exponentially with bandwidth and cardinality of antenna branches. As an example, for a 5G mid-band carrier with 30 kHz subcarrier spacing, 100 MHz bandwidth, and 64 transmit and receive antennas, about 85% of the processing demand of the RAN protocol layers comes from the PHY layer (including beamforming) [4].

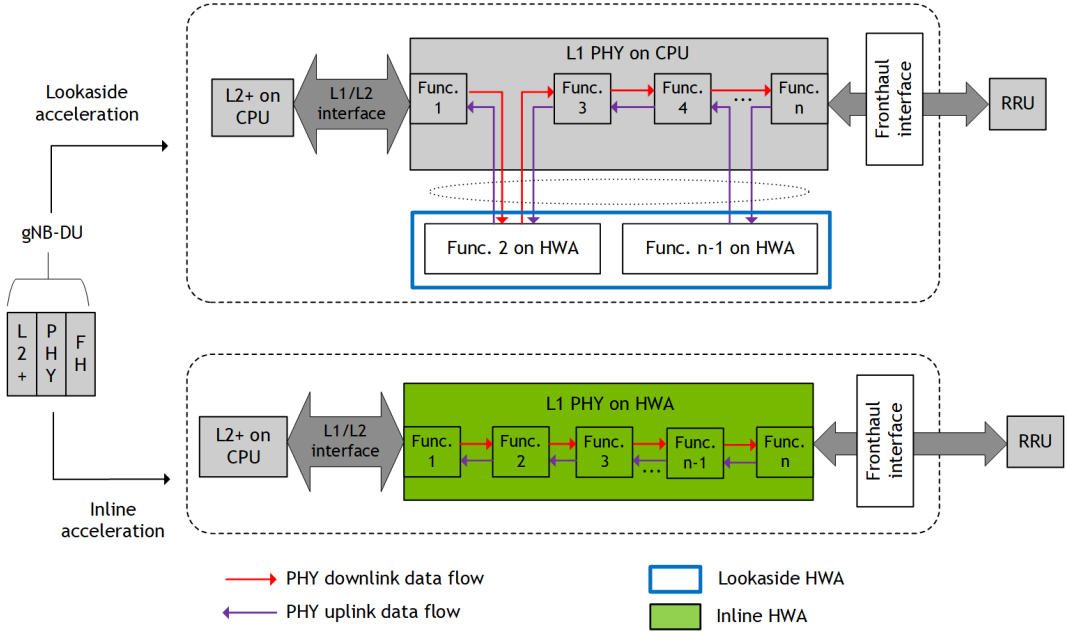

**Figure 2 An illustration of gNB-DU layer 1 (PHY) acceleration in lookaside vs. inline mode**

The rising demand for L1 processing in a 5G RAN calls for accelerated computing, which often cannot be fulfilled by GPPs alone. A purely central processing unit (CPU) based 5G RAN platform can drive cost and power consumption levels beyond practical limits. At that point, hardware acceleration becomes an inevitable choice, where computationally complex PHY layer signal processing is offloaded from host CPU to HWA. In general, HWAs are specialized hardware implementations, capable of processing compute-heavy workloads more efficiently than GPPs. The choice of HWAs ranges from fully programmable processors like GPU and DPU, to heavily customized, largely fixed function accelerators with varying degrees of programmability like FPGA, DSP, ASIC, and SoC. Using HWA, the PHY layer is implemented either as a purely hardware solution (i.e., running on HWA), or as a hybrid 'software plus hardware' solution (i.e., running partly on CPU and partly on HWA), depending on the specific use case and to what extent accelerated compute is warranted.

In general, the mode of hardware acceleration can be either *lookaside* or *inline* [12], as illustrated in Figure 2. In lookaside mode, the host CPU invokes data processing offload to an HWA and receives the result back from the HWA once the processing is complete. Since the 4G LTE era, traditional RAN accelerations for PHY layer have primarily been implemented in this mode, e.g., forward error correction (FEC) for physical control and shared channels. Channel coding is one of the most daunting signal processing tasks in L1. An FEC accelerator enables faster and more efficient processing of various channel encoding/decoding algorithms than CPU [8].

One of the limitations of lookaside acceleration is the back-and-forth data transfer between the host CPU and the HWA. If multiple non-contiguous functions are accelerated in lookaside mode as shown in Fig. 2, the overhead of repeated host-to-device data transfer and the resulting double data rate (DDR) bandwidth consumption go up significantly.

With the advent of massive MIMO antenna arrays, in conjunction with new and wider 5G mid/high frequency bands and shorter TTIs, the computing demand for 5G L1 is increasing at an exponential rate, pushing the limit of lookaside acceleration to the edge. At this juncture, inline acceleration mode has emerged as a promising solution for mitigating the massive data transfer bottleneck of lookaside approach [12]. The core principle of inline PHY acceleration is to offload the entire L1 processing pipeline to HWA, and upon completion of the processing, to send the output data directly from HWA to the next destination (i.e., the RRU) instead of routing via CPU. Thus, inline PHY acceleration avoids the redundant ping-pong data path that throttles the interface between CPU and HWA. Enabling fast input/output (I/O) processing (e.g., by using smart network interface card (SmartNIC) and direct memory access (DMA) technology [13]) is essential in unfolding the full potential of inline acceleration.

As 5G RAN is undergoing a momentous transformation towards open and multi-vendor architecture, it is pertinent to rethink hardware acceleration from a different perspective - how to make this diverse set of acceleration options (including acceleration modes, accelerated functions, and accelerating devices) compatible with various RAN applications, potentially coming from different software vendors. To that end, O-RAN Alliance is developing a framework to abstract acceleration differences from the network applications through acceleration abstraction layer (AAL), which is detailed in the next section.

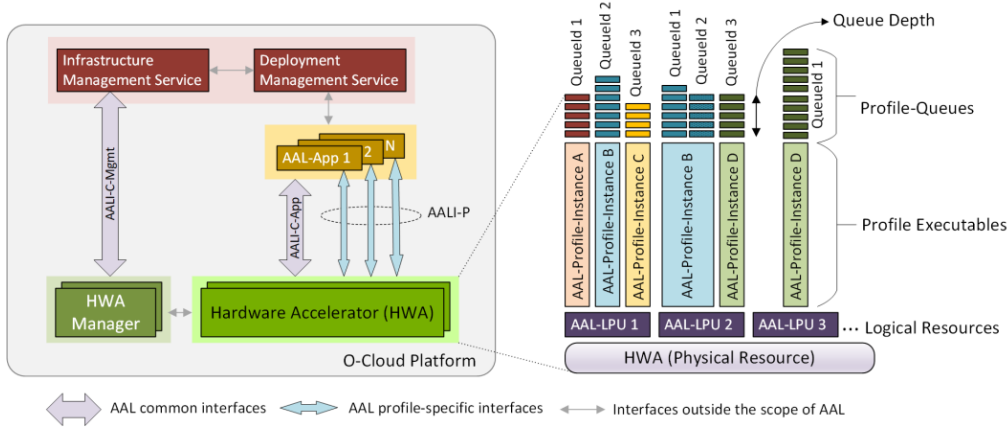

**Figure 3 High level architecture of AAL and the three-tier abstraction of underlying HWA**

## IV. ACCELERATION ABSTRACTION IN OPEN RAN

One of the fundamental tenets of open RAN is to leverage cloud-native principles in facilitating RAN application portability across various hardware platforms. Although traditional vRAN replaces proprietary hardware platforms with COTS-based servers, the VNFs/CNFs still remain dependent on the specific HWA type on which the software runs, with tightly coupled interfaces in between them. O-RAN aims to abstract the RAN software from underlying HWAs through AAL interface (AALI), so that application vendors can flexibly execute their software on any type of HWA or migrate their applications from one HWA to another in a plug-and-play way, without requiring custom integration.

### A. AAL Architecture

The notion of AAL evolves around two key aspects - 1) *what-to-accelerate* and 2) *how-to-accelerate*.

*What-to-accelerate:* AAL introduces the concept of AAL-profile, which specifies a set of function(s) to be accelerated by an HWA on behalf of an AAL application (AAL-App). Through AALI, an AAL-App offloads AAL-profile specific workload to an HWA for accelerated processing.

*How-to-accelerate:* An AAL-App accesses an underlying HWA through a three-tier abstraction - AAL logical processing unit (AAL-LPU), AAL-profile-instance and AAL-profile-queue as depicted in Figure 3. HWA resources are logically represented to an AAL-App through AAL-LPU(s). An AAL-App communicates with an AAL-LPU in the same way irrespective of whether the underlying HWA resources are hard-partitioned (e.g., in a multi-instance GPU (MIG)) or soft-partitioned (e.g., in a single-root input/output virtualization (SRIOV) capable FPGA), and thereby abstracted from specific HWA type/vendor. Within each AAL-LPU, an AAL-App can accelerate its workloads using one or more AAL-profile-instances, which are the executables of an AAL-profile. Each AAL-profile-instance can further expose one or more AAL-profile-queues (each associated with a QueueId), using which an AAL-App can group AAL-profile specific operations together and schedule jobs with different priorities in different queues, with each queue having a certain depth for serialized workload consumption. Internally, AAL-profile-queues may access distinct resources of an AAL-LPU (e.g., compute, I/O), but these implementation specific details are obscured from the AAL-App through the AALI.

Figure 3 depicts a high level architecture for AAL, which is constituted of two key interfaces: AAL-common-interface (AALI-C) and AAL-profile-specific-interface (AALI-P). Through AALI-C, AAL-App can communicate with underlying HWA in a profile-independent way. On the other hand, AALI-P is constituted of a set of profile-specific application programming interfaces (APIs), using which AAL-App establishes profile-specific communication with HWA. AALI-P varies from one AAL-profile to another. Within AALI-C, there are two categories, viz. AALI-C-management (AALI-C-Mgmt) and AALI-C-application (AALI-C-App). While AALI-C-Mgmt supports a set of APIs for common administrative operations, actions, and events associated with the management of the HWA, AALI-C-App provides APIs for operations between AAL-App and HWA. AALI-C APIs are currently under development by O-RAN, while AALI-P APIs are maturing primarily for two AAL-profiles: FEC in lookaside mode and high-PHY in inline mode. The latter one is essentially the acceleration profile for O-DU PHY layer or L1.

From an AAL-app point of view, AALI is defined based on a set of generic features for AALI-C-App interface. These generic interface principles include API extensibility, support for interrupt and poll modes, capability to discover and configure AAL-LPUs, options for different acceleration offload architectures (e.g., lookaside, inline or hybrid), support for multi-threading environment, capability of handling versatile ranges of acceleration payload, separation of control and user plane dataflows, and abstraction of different transport mechanisms between the AAL-App and HWA. On the other hand, from an open cloud (O-cloud) infrastructure point of view, the AALI provides a consistent interface through a set of AALI-C-Mgmt APIs, built upon the generic principles of supporting administrative operations, actions, and events related to HWA which is part of an O-cloud platform hardware. Operations supported through AALI-C-Mgmt APIs include discovery, lifecycle management, performance monitoring, configuration, updates/upgrades and error handling. Together, AALI-C-App and AALI-C-Mgmt APIs constitute an abstraction layer between various AAL-Apps running on an O-cloud platform and the platform's underlying HWA resources. Thus, AALI enables hardware-software decoupling and offers a generic way of using and managing different types of HWAs within the infrastructure of a cloud-native open RAN.

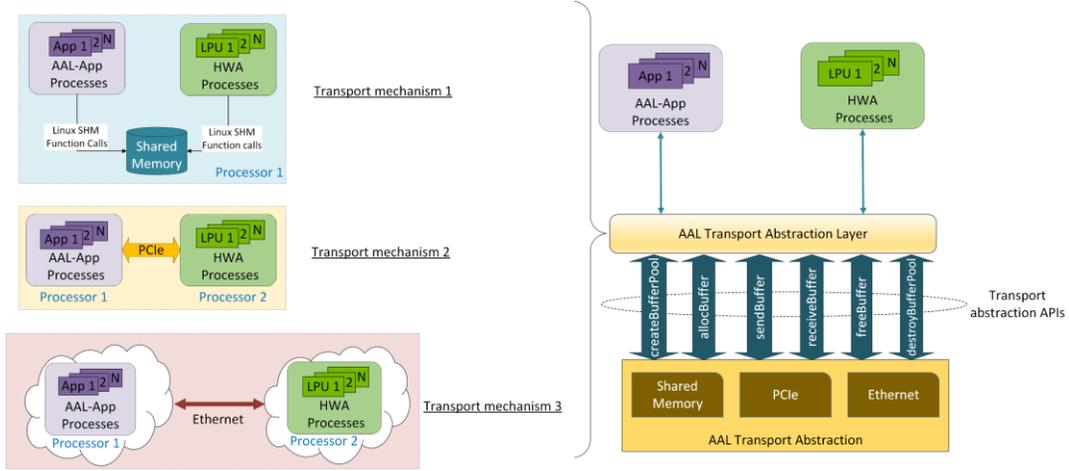

**Figure 4 Illustration of transport generalization in AAL transport abstraction framework**

### B. AAL Transport Abstraction

One of the most important features of AALI-C-App is transport abstraction between the AAL-App and the underlying HWA [14]. In general, this transport can be of different types, for example, based on shared memory (SHM), via peripheral component interconnect express (PCIe), or over Ethernet. AALI-C-App abstracts these various mechanisms from the AAL-Apps through a set of generic transport APIs, constituting a transport abstraction framework as depicted in Figure 4. The transport abstraction framework supports a set of operations (and associated APIs) for the AAL-Apps to create and manage buffers in a generic and consistent way without dictating implementation specific memory management of HWAs. The set of transport APIs generalizes common buffer management operations like buffer pool creation, buffer allocation, sending and reception of buffers to/from HWA, freeing pre-allocated buffers, and terminating buffer pools.

The AAL transport abstraction framework is designed to support flexible HWA implementations, while providing a set of generic APIs to cater to diverse use cases. As one example, buffer sending and reception APIs are supported for both synchronous and asynchronous modes of operations. When a synchronous API is invoked by an AAL-App, the underlying HWA (i.e., the API implementor) blocks the application thread until the corresponding operation is complete and upon completion, the HWA returns the status of the operation execution (success/failure) to the AAL-App. On the other hand, for an asynchronous API, the operation is executed in a non-blocking mode, i.e., the application thread invoking the API call is immediately returned, before the execution completion of the send/receive operation by the HWA. Upon completion, the HWA optionally returns the status of the operation to the AAL-App if it had preregistered for a callback.

The other example of flexible transport is related to the options supported for buffer ownership transfer. The API related to sending buffers supports a mechanism to indicate whether the ownership of the enqueue buffers is transferred from the AAL-App to the HWA with the API invocation, or retained by the AAL-App. In cases where the AAL-App needs to hold onto an enqueued buffer for buffer-recycling (i.e., reusing the same buffer for later use) or for supporting special use cases like RLC acknowledgement mode (AM), the AAL-App retains the buffer ownership and such buffers can only be freed by explicit invocation of a buffer freeing API later by the AAL-App. When the AAL-App does not need to retain an enqueued buffer, it transfers the buffer ownership to the HWA with the send API invocation, in which case the HWA autonomously releases the buffer upon successful completion of the send operation, without requiring an explicit buffer free API invocation by the AAL-App.

Finally, the buffer receive API is defined with the flexibility of corresponding dequeue buffer allocation being managed either by the AAL-App or the HWA. When the AAL-App allocates dequeue buffers, it indicates so in the buffer receive API and the HWA transfers received data into these buffers. Without such indication from the AAL-App, the HWA allocates dequeue buffers internally and returns those buffers to the AAL-App for fetching received data. This way, different HWA implementation options can be supported by the same, unified transport abstraction framework.

AAL transport abstraction APIs, in conjunction with AALI-P APIs for high-PHY profile, pave the way for multi-vendor, high performance O-DU solutions, enabling integration of vendor A's DU stack with layer 2 (L2) and above with vendor B's accelerated L1 stack, in a truly plug-and-play way.

To summarize, hardware/software disaggregation, in conjunction with acceleration and transport abstraction enabled by O-RAN opens up the possibility of realizing multi-vendor, cloud-native, interoperable and high-performance RAN solutions by leveraging the best-of-the-breed capabilities of HWAs, which is the hallmark of a successful adoption pathway for open RAN. One such multi-vendor, high performance open RAN solution with GPU accelerated L1 is described in detail in the following section.

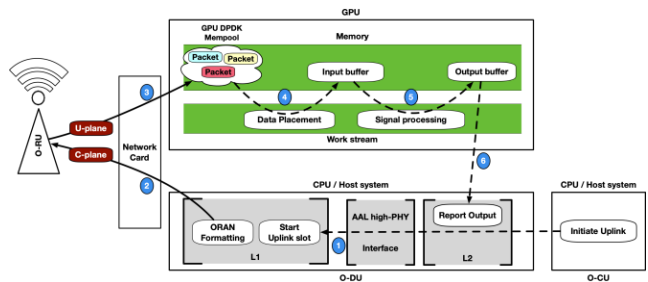

Figure 5 Illustration of uplink data and control flows

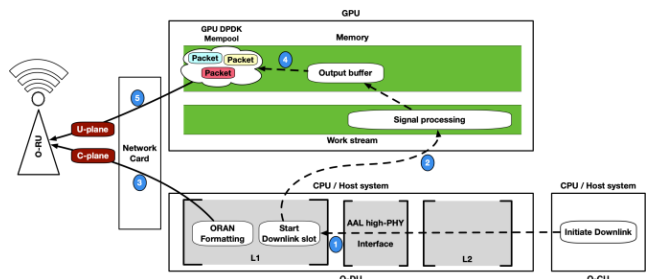

Figure 6 Illustration of downlink data and control flows

## V. USE CASE: INLINE HARDWARE ACCELERATION FOR L1

Selection of specific HWA and acceleration mode in open RAN deployments primarily depends on three key factors - programmability, compute efficiency, and power consumption. In general, a fully programmable, software-defined solution like GPU accelerated inline high-PHY [10] is one of the most versatile options, offering both high performance as well as agility and scalability as per MNOs' compute needs. In this section, we will deep dive into a GPU-based practical implementation of inline hardware acceleration for AAL high-PHY profile (i.e., O-DU L1). For an efficient implementation, separate design aspects are considered depending on the direction of the data flow, i.e., uplink (data ingress into HWA) versus downlink (data egress from HWA).

### A. Uplink

Accelerating L1 signal processing in the uplink direction can be considered as an inline packet processing application, where packets with in-phase/quadrature-phase (IQ) samples arriving at the network interface card (NIC) from different O-RUs are received, validated, and reordered in real-time, and provided as an input to the signal processing pipeline to be accelerated on the HWA. To sustain high network throughput, it is crucial to enable fast I/O processing and fronthaul packet ingress through efficient orchestration of message flows in between the three components of O-DU, viz. CPU, HWA, and NIC.

In this context, there are two separate paths of message flow to be considered: 1) data path and 2) control path. In the data path, user plane (U-plane) packets from O-RU are directly received via the NIC into the HWA memory using DMA over PCIe. In the control path, the CPU, after sending control plane (C-plane) packets to O-RU, queries the NIC about new incoming U-plane packets, and once packets arrive in the HWA memory, triggers the uplink signal processing pipeline.

In a GPU-based implementation, these paths can be realized in different ways. One approach is to receive and accumulate packets in the GPU memory for an entire over-the-air (OTA) time slot, before launching the GPU processing to rebuild the full payload from received packets and trigger the uplink signal processing pipeline. This sequential packet processing is suboptimal from the latency point of view. An alternate, more performant approach is to have real-time communication among CPU, GPU, and NIC, where as soon as a subset of packets is accumulated, the GPU starts validation and reordering of those packets while more packets continue to be coming from the network – i.e., parallelizing the processes of packet reception and packet ordering. That way, at the end of every OTA slot, the uplink payload is already built and validated at the GPU memory so that the uplink signal processing pipeline can be triggered immediately.

To implement the second approach, NVIDIA has designed a solution generic enough to be adopted by any GPU-based inline HWA to receive and process packets in real-time through the data plane development kit (DPDK) framework. To that end, a new open-sourced library called GPUdev [15] is introduced in DPDK to enhance communications between CPU, NIC, and GPU. In particular, GPUDev enables detection of GPU as a PCIe device within the DPDK framework, enables easy reception (or sending) of packets directly into (or from) the GPU memory without CPU memory staging copies using GPUDirect remote DMA (RDMA) [15], and offers a highly efficient, low-latency communication mechanism between the CPU and the GPU.

Figure 5 illustrates the uplink data and control flows with the design considerations stated above. The uplink processing is constituted of the following steps: 1) higher layer (L2) initiates uplink slot processing by sending configuration request using AAL high-PHY APIs, 2) L1 control path encapsulates C-plane messages with O-RAN formatting and sends to O-RU via NIC, 3) O-RU sends U-plane packets over fronthaul which are directly stored in the GPU memory, 4) out-of-order received packets are reordered through data placement action and placed into contiguous input memory buffer, 5) once packet ordering is complete, uplink signal processing is initiated on GPU, and 6) in post processing, the output is transferred from the GPU output buffer to the higher layer (L2) using AAL transport abstraction APIs.

### B. Downlink

Accelerating L1 signal processing in the downlink direction requires the HWA to process transport blocks (TBs) received from higher layers and once done, to send the IQ samples over the fronthaul to the O-RU. Similar to the uplink, there are two message flow paths in the downlink as well, viz. 1) control path and 2) data path. Since the control path processing is much less compute-intensive than data path processing, one approach of implementing downlink flows would be to keep the control path processing in the CPU, while offloading the data path processing to the HWA. In that approach, the control path processing of L1 running on the CPU receives configurations for the processing pipeline of a subsequent downlink slot and triggers the signal processing on the HWA which will store the processing outcome (e.g., IQ samples) in its output buffer.

In the context of a GPU-based implementation, while the processing of TBs can be offloaded to the GPU, the ethernet packetization of both C-plane and U-plane messages can be done by CPU through DPDK functions. Especially, for the U-plane packets the DPDK external buffer feature can be used, which allows a packet to be composed by fetching data from different memory locations. Since the U-plane payload resides in the GPU memory at the end of the downlink signal processing, each U-plane packet is formed by two different memory areas, one pointing to a CPU memory location where CPU prepares and stores the packet headers (including ethernet and O-RAN specific headers) and a second one pointing to the GPU memory buffer where the packet payload is stored.

Figure 6 illustrates the downlink data and control flows with the design considerations stated above. The downlink processing is constituted of the following steps–: 1) higher layer (L2) initiates downlink slot processing by sending configuration request using AAL high-PHY APIs, 2) TB is transferred from the host system (running L2) to GPU using AAL transport abstraction APIs and downlink processing pipeline is triggered on GPU, 3) L1 control path encapsulates C-plane messages with O-RAN formatting and sends to O-RU via NIC, 4) in post downlink processing, the generated IQ samples in the output buffer are stored into GPU memory pool as packet payloads and 5) directly sent to NIC for ethernet encapsulation before being pushed to O-RU over fronthaul.

Inline acceleration implementation of AAL high-PHY described above and the corresponding dataflows are primarily 'CPU-centric', which means that the CPU is in the critical path acting as a man-in-the-middle agent between the GPU and the NIC to orchestrate and coordinate tasks between these two devices. A possible optimization would be to move to a 'GPU-centric' approach where the GPU can directly interact with the NIC to send or receive packets without waiting for the CPU's intervention. This way, the GPU can independently work on data processing and network activity, reducing CPU responsibilities. An in-depth discussion of this optimized approach is out of the scope of this paper and will be addressed in the future.

## VI. CONCLUSION AND FUTURE OUTLOOK

We are at an inflection point where we see a paradigm shift in the RAN evolution towards open, interoperable, intelligent, software-defined, and cloud-native solutions. HWAs are essential for enabling high-performance and energy-efficient open RAN. This article has provided a holistic overview of the state of the art in hardware acceleration for open RAN. We anticipate that hardware acceleration will play an even more critical role in meeting the demands of the sixth generation (6G) wireless networks.

The field of hardware acceleration for open RAN is an active area of research. We conclude by pointing out some important avenues for future work.

- *Beyond PHY layer processing*: Currently hardware acceleration is mainly used for the PHY layer processing in RAN. One key direction is to explore the role of hardware acceleration beyond PHY, such as for scheduling in L2 or artificial intelligence (AI) workloads on RAN intelligent controller (RIC) platforms.
- *Multitenant cloud RAN*: Pooling HWAs into a cloud-native environment allows for dynamically orchestrating resources between the RAN processing and off-peak workloads. Examples of RAN off-peak workloads are AI workloads in urban areas such as drive mapping and offline video analytics. It is of interest to investigate how to utilize HWAs efficiently and effectively for RAN in the cloud.
- *Standardization*: As the use of HWAs becomes more widespread in open RAN deployments, there will be a growing need for further standardization, calling for new APIs, interfaces, protocols, testing and certification. Continuous standardization will be the key to enabling interoperability and promoting innovative use of HWAs in open RAN evolution towards 6G.

BIOGRAPHIES

**Lopamudra Kundu** is a Senior Standards Engineer at NVIDIA, leading O-RAN Alliance delegation and driving open RAN ecosystem engagement.

**Xingqin Lin** is a Senior 3GPP Standards Engineer at NVIDIA, leading 3GPP standardization and conducting work at the intersection of 5G/6G and AI.

**Elena Agostini** is a Senior Software Engineer at NVIDIA, currently working on NVIDIA GPUDirect Technologies and DOCA framework. She is also a contributor to DPDK.

**Vikrama Ditya** is the Director of 5G RAN Compute & Standards at NVIDIA, leading a team addressing GPU accelerated software, reference system, benchmarking and algorithm design/development for 5G vRAN.